\documentclass[titlepage,12pt]{article}
\usepackage{amssymb}
\usepackage{amsfonts}
\usepackage{amsmath}
\usepackage[dvips]{graphicx}
\usepackage{subfigure}
\usepackage[font=small,format=plain,labelfont=bf,up,textfont=it,up]{caption}
\usepackage{appendix}

\begin{document}

\title{Fermions in a mixed vector-scalar double-step potential via
continuous chiral transformation\thanks{%
To appear in The European Physical Journal Plus }}
\date{}
\author{W.M. Castilho\thanks{%
E-mail address: castilho.w@gmail.com (W.M. Castilho)} and A.S. de Castro%
\thanks{%
E-mail address: castro@pq.cnpq.br (A.S. de Castro)} \\
%EndAName
\\
UNESP - Campus de Guaratinguet\'{a}\\
Departamento de F\'{\i}sica e Qu\'{\i}mica\\
12516-410 Guaratinguet\'{a} SP - Brazil }
\date{}
\maketitle

\begin{abstract}
The behaviour of fermions in the background of a double-step potential is
analyzed with a general mixing of scalar and vector couplings via continuous
chiral-conjugation transformation. Provided the vector coupling does not
exceed the scalar coupling, a Sturm-Liouville approaching for the
double-step potential shows that the transmission coefficient exhibits
oscillations and that a finite set of intrinsically relativistic bound-state
solutions might appear as poles of the transmission amplitude in a strong
coupling regime. An isolated bound-state solution resulting from coupled
first-order equations might also come into sight. It is also shown that all
those possible bound solutions disappear asymptotically as one approaches
the conditions for the realization of the so-called spin and pseudospin
symmetries in a four-dimensional space-time. Furthermore, we show that due
to the additional mass acquired by the fermion from the scalar background
the high localization of the fermion in an extreme relativistic regime does
not violate the Heisenberg uncertainty principle.
\end{abstract}

\section{Introduction}

The Dirac Hamiltonian with a mixing of scalar potential and time component
of vector potential in a four-dimensional space-time is invariant under an
SU(2) algebra when the difference between the potentials, or their sum, is a
constant \cite{br}. The near realization of these symmetries may explain
degeneracies in some heavy meson spectra (spin symmetry) \cite{pag}-\cite%
{gin} or in single-particle energy levels in nuclei (pseudospin symmetry)
\cite{gin}-\cite{ps}. When these symmetries are realized, the energy
spectrum does not depend on the spinorial structure, being identical to the
spectrum of a spinless particle \cite{spin0}. Despite the absence of spin
effects in 1+1 dimensions, many attributes of the spin and pseudospin
symmetries in four dimensions are preserved.

In a pioneering work, Jackiw and Rebbi \cite{jr} have shown that massless
fermions coupled to scalar fields with kink-like profiles in 1+1 dimensions
develops quantum states with fractional fermion number due to the zero-mode
solution. This phenomenon has been seen in certain polymers such as
polyacetylene \cite{kink}. Later, charge fractionization into irrational
numbers was shown to emerge in a model without charge-conjugation symmetry
\cite{gol}. Charge values with irrational numbers are present in a certain
discrete model of diatomic polymers \cite{ric} which is related to
charge-conjugation-invariance violation in the continuum limit \cite{jac},
realizing the charge fractionization envisioned in Ref. \cite{gol}.

Recently the complete set of solutions for the kink-like scalar potential
behaving like $\mathrm{\tanh }\,x\!/\!\lambda $ has been considered for
massless fermions in Ref. \cite{char1}, and for massive fermions in Ref.
\cite{char2}. Kink-like profiles violating the charge-conjugation symmetry
by addition of a time component of a vector potential has also been
considered in the literature. The complete set of solutions for massive
fermions under the influence of a kink-like scalar potential added by the
time component of a vector potential with the same functional form was
considered in Refs. \cite{aop1} and \cite{aop2}, in Ref. \cite{aop1} for the
background field behaving like $\mathrm{sgn}\,x$, and in Ref. \cite{aop2}
for the background behaving like $\mathrm{\tanh }\,x\!/\!\lambda $. In Refs.
\cite{aop1} and \cite{aop2}, it has been shown that the Dirac equation with
a scalar potential plus a time component of vector potential of the same
functional form is manageable if the vector coupling does not exceed the
scalar coupling, and that the bound states for mixed scalar-vector
potentials with the kink-like profiles $\mathrm{sgn}\,x$ and $\mathrm{\tanh }%
\,x\!/\!\lambda $ are intrinsically relativistic solutions. Furthermore, it
has been shown that the fermion can be confined in a highly localized region
of space under a very strong field without any chance for spontaneous pair
production related to Klein's paradox. In a more recent work \cite{jpcs}, it
has been shown that the existence of such intrinsically relativistic bound
states is ensured for any mixed scalar-vector potential with a kink-like
profile.

Elsewhere, it has been shown that the double-step potential can furnish
intrinsically relativistic bound states when one considers a pseudoscalar
coupling in the Dirac equation \cite{pseu} or a nonminimal vector coupling
in the Duffin-Kemmer-Petiau equation for spinless particles \cite{oli} and
spin-1 particles \cite{oli2}. However, the double-step potential does not
seem to have received any attention regarding scalar and vector couplings in
the Dirac equation. In what follows the mixing formalism developed in \cite%
{aop1}-\cite{aop2} is addressed to the double-step potential. We shall
present a fairly complete account of the problem and show that, in contrast
to the case of a sign step potential of Ref. \cite{aop1} and alike the
smooth step potential of Ref. \cite{aop2}, aside from the isolated solution
the spectrum might consist of a finite set of bound-state solutions under a
strong-coupling regime. Furthermore, we shall show that the fermion can be
trapped in a highly localized region under an extreme relativistic regime
and that this high localization maintains the single-particle interpretation
of the Dirac theory because the fermion acquires an additional mass coming
from its interaction with the scalar-field background. A distinctive aspect
of this problem is that the transmission coefficient exhibits oscillations.
The limit where the double-step potential becomes the sign potential is also
considered. We begin by reviewing those results of Refs. \cite{aop1}, \cite%
{aop2} and \cite{jpcs} which are directly relevant for the present work.

\section{Mixed scalar-vector interactions}

Consider the Lagrangian density for a massive fermion%
\begin{equation}
L=\bar{\Psi}\left( i\hbar c\gamma ^{\mu }\partial _{\mu }-Imc^{2}-V\right)
\Psi  \label{d1}
\end{equation}%
where $\hbar $ is the constant of Planck, $c$ is the velocity of light, $I$
is the unit matrix, $m$ is the mass of the free fermion and the square
matrices $\gamma ^{\mu }$ satisfy the algebra $\{\gamma ^{\mu },\gamma ^{\nu
}\}=2Ig^{\mu \nu }$. The spinor adjoint to $\Psi $ is defined by $\bar{\Psi}%
=\Psi ^{\dagger }\gamma ^{0}$. For vector and scalar interactions the matrix
potential is written as $V=\gamma ^{\mu }A_{\mu }+IV_{s}$. Requiring $\left(
\gamma ^{\mu }\right) ^{\dag }=\gamma ^{0}\gamma ^{\mu }\gamma ^{0}$, one
finds the continuity equation $\partial _{\mu }J^{\mu }=0$, where the
conserved current is $J^{\mu }=c\bar{\Psi}\gamma ^{\mu }\Psi $. Eq. (\ref{d1}%
) leads to the Hamiltonian form for the Dirac equation $i\hbar \partial \Psi
/\partial t=H\Psi $. In 1+1 dimensions $\Psi $ is a 2$\times $1 matrix, the
metric tensor is $g^{\mu \nu }=$ diag$\left( 1,-1\right) $ and the
Hamiltonian is given by%
\begin{equation}
H=\gamma ^{5}c\left( p_{1}+\frac{A_{1}}{c}\right) +IA_{0}+\gamma ^{0}\left(
mc^{2}+V_{s}\right)
\end{equation}%
where $\gamma ^{5}=\gamma ^{0}\gamma ^{1}$. Assuming time-independent
potentials, one can write $\Psi \left( x,t\right) =\psi \left( x\right) \exp
\left( -iEt/\hbar \right) $ in such a way that the time-independent Dirac
equation becomes $H\psi =E\psi $. Meanwhile $J^{\mu }=c\overline{\psi }%
\gamma ^{\mu }\psi $ is time independent and $J^{1}$ is uniform. From now
on, we make $A_{1}=0$ and use an explicit representation for the 2$\times $2
matrices $\gamma $ as $\gamma ^{0}=\sigma _{3}$ and $\gamma ^{1}=i\sigma
_{2} $. Here, $\sigma _{2}$ and $\sigma _{3}$ stand for the Pauli matrices.

The charge-conjugation operation is accomplished by the transformation $\psi
\rightarrow \sigma _{1}\psi ^{\ast }$ followed by $A_{0}\rightarrow -A_{0}$,
$V_{s}\rightarrow V_{s}$ and $E\rightarrow -E$ \cite{prc}. The
chiral-conjugation operation $\psi \rightarrow \gamma ^{5}\psi $ (according
to Ref. \cite{wat}) is followed by the changes of the signs of $V_{s}$ and $%
m,$ but not of $A_{0}$ and $E$ \cite{prc}. One sees that the
charge-conjugation and the chiral-conjugation operations interchange the
roles of the upper and lower components of the Dirac spinor. The continuous
chiral transformation (see, e.g., \cite{tou}) is induced by the unitary
operator
\begin{equation}
U(\theta )=\exp \left( -\frac{\theta }{2}i\gamma ^{5}\right)  \label{2}
\end{equation}%
\noindent where $\theta $ is a real quantity such that $0\leq \theta \leq
\pi $. It allows one to write
\begin{equation}
h\phi =E\phi ,\quad \phi =U\psi ,\quad h=UHU^{-1}  \label{22}
\end{equation}%
with
\begin{equation}
h=\sigma _{1}cp_{1}+IA_{0}+\sigma _{3}\left( mc^{2}+V_{s}\right) \cos \theta
-\sigma _{2}\left( mc^{2}+V_{s}\right) \sin \theta  \label{333333}
\end{equation}%
With $A_{0}$ and $V_{s}$ related by
\begin{equation}
A_{0}=V_{s}\cos \theta  \label{5}
\end{equation}%
\noindent and eliminating $A_{0}$ in favor of $V_{s}$, one can rewrite the
Dirac equation in terms of the upper ($\phi _{+}$) and lower ($\phi _{-}$)
components of $\phi $ as
\begin{subequations}
\begin{eqnarray}
\hbar c\frac{d\phi _{+}}{dx}+\left( mc^{2}+V_{s}\right) \sin \theta \,\phi
_{+} &=&i\left( E+mc^{2}\cos \theta \right) \phi _{-}  \label{6a} \\
&&  \notag \\
\hbar c\frac{d\phi _{-}}{dx}-\left( mc^{2}+V_{s}\right) \sin \theta \,\phi
_{-} &=&i\left[ E-\left( mc^{2}+2V_{s}\right) \cos \theta \right] \phi _{+}
\label{6b}
\end{eqnarray}

For $E\neq -mc^{2}\cos \theta $, \noindent one finds
\end{subequations}
\begin{equation}
J^{1}=\frac{2\hbar c^{2}}{E+mc^{2}\cos \theta }\,\text{Im}\left( \phi
_{+}^{\ast }\frac{d\phi _{+}}{dx}\right)
\end{equation}%
Furthermore,
\begin{equation}
-\frac{\hbar ^{2}}{2}\frac{d^{2}\phi _{+}}{dx^{2}}+V_{\mathtt{eff}}\,\phi
_{+}=E_{\mathtt{eff}}\,\phi _{+}  \label{8}
\end{equation}%
with%
\begin{equation}
V_{\mathtt{eff}}=\frac{\sin ^{2}\theta }{2c^{2}}V_{s}^{2}+\frac{mc^{2}+E\cos
\theta }{c^{2}}V_{s}-\frac{\hbar \sin \theta }{2c}\frac{dV_{s}}{dx}
\label{v}
\end{equation}%
and%
\begin{equation}
E_{\mathtt{eff}}=\frac{E^{2}-m^{2}c^{4}}{2c^{2}}  \label{e}
\end{equation}%
\noindent In this way one can solve the Dirac problem for determining the
possible discrete or continuous eigenvalues of the system by referring to
the solution of the related Schr\"{o}dinger problem because $\phi _{+}$ is a
square-integrable function.

Defining%
\begin{equation}
v\left( x\right) =\int^{x}dy\,V_{s}\left( y\right)
\end{equation}%
the solutions for (\ref{6a}) and (\ref{6b}) with $E=-mc^{2}\cos \theta $ are
\begin{subequations}
\begin{eqnarray}
\phi _{+} &=&N_{+}  \label{ii1a} \\
&&  \notag \\
\phi _{-} &=&N_{-}-2\frac{i}{\hbar c}N_{+}\left[ mc^{2}x+v\left( x\right) %
\right] \cos \theta  \label{ii1b}
\end{eqnarray}%
for $\sin \theta =0$, and
\end{subequations}
\begin{subequations}
\begin{eqnarray}
\phi _{+} &=&N_{+}\exp \left\{ -\frac{\sin \theta }{\hbar c}\left[
mc^{2}x+v\left( x\right) \right] \right\}  \label{ii2a} \\
&&  \notag \\
\phi _{-} &=&N_{-}\exp \left\{ +\frac{\sin \theta }{\hbar c}\left[
mc^{2}x+v\left( x\right) \right] \right\} +i\phi _{+}\cot \theta
\label{ii2b}
\end{eqnarray}%
for $\sin \theta \neq 0$. \noindent $N_{+}$ and $N_{-}$ are normalization
constants. Note that these solutions isolated from the Sturm-Liouville
problem can not describe scattering states and $J^{1}=2c\text{Re}\left(
N_{+}^{\ast }N_{-}\right) $. A bound-state solution demands $N_{+}=0$ or $%
N_{-}=0$, because $\phi _{+}$ and $\phi _{-}$ are square-integrable
functions. There is no bound-state solution for $\sin \theta =0$, and for $%
\sin \theta \neq 0$ the existence of a bound state solution depends on the
asymptotic behaviour of $v(x)$ \cite{hot}.

\section{Kink potentials}

Now we consider a kink-like potential with the asymptotic behaviour $%
V_{s}(x)\rightarrow \pm v_{0}$ as $x\rightarrow \pm \infty $, with $v_{0}=$
constant.

We turn our attention to scattering states for fermions with $E\neq
-mc^{2}\cos \theta $ coming from the left. Then, $\phi $ for $x\rightarrow
-\infty $ describes an incident wave moving to the right and a reflected
wave moving to the left, and $\phi $ for $x\rightarrow +\infty $ describes a
transmitted wave moving to the right or an evanescent wave. The upper
component for scattering states is written as
\end{subequations}
\begin{equation}
\phi _{+}=\left\{
\begin{array}{cc}
A_{+}e^{+ik_{-}x}+A_{-}e^{-ik_{-}x}, & \text{for\quad }x\rightarrow -\infty
\\
&  \\
B_{\pm }e^{\pm ik_{+}x}, & \text{for\quad }x\rightarrow +\infty%
\end{array}%
\right.  \label{phi}
\end{equation}%
where
\begin{equation}
\hbar ck_{\pm }=\sqrt{\left( E\mp v_{0}\cos \theta \right) ^{2}-\left(
mc^{2}\pm v_{0}\right) ^{2}}
\end{equation}%
Therefore,%
\begin{equation}
J_{\gtrless }^{1}\left( -\infty \right) =\frac{2\hbar c^{2}k_{-}}{%
E+mc^{2}\cos \theta }\left( |A_{\pm }|^{2}-|A_{\mp }|^{2}\right) ,\text{%
\quad for\quad }E\gtrless -mc^{2}\cos \theta
\end{equation}%
and
\begin{equation}
J_{\gtrless }^{1}\left( +\infty \right) =\pm \,\frac{2\hbar c^{2}\text{Re}%
\,k_{+}}{E+mc^{2}\cos \theta }\,|B_{\pm }|^{2},\text{\quad for\quad }%
E\gtrless -mc^{2}\cos \theta
\end{equation}%
If $E>-mc^{2}\cos \theta $, then $A_{+}e^{+ik_{-}x}$ ($A_{-}e^{-ik_{-}x}$)
will describe the incident (reflected) wave, and $B_{-}=0$. On the other
hand, if $E<-mc^{2}\cos \theta $, then $A_{-}e^{-ik_{-}x}$ ($%
A_{+}e^{+ik_{-}x}$) will describe the incident (reflected) wave, and $%
B_{+}=0 $. Therefore, the reflection and transmission amplitudes are given by%
\begin{equation}
r_{\gtrless }=\frac{A_{\mp }}{A_{\pm }},\text{\quad }t_{\gtrless }=\frac{%
B_{\pm }}{A_{\pm }},\text{\quad for\quad }E\gtrless -mc^{2}\cos \theta
\label{t}
\end{equation}
The $x$-independent space component of the current allows us to define the
reflection and transmission coefficients as%
\begin{equation}
R_{\gtrless }=\frac{|A_{\mp }|^{2}}{|A_{\pm }|^{2}},\text{\quad }T_{\gtrless
}=\frac{\text{Re}\,k_{+}}{k_{-}}\frac{|B_{\pm }|^{2}}{|A_{\pm }|^{2}},\text{%
\quad for\quad }E\gtrless -mc^{2}\cos \theta  \label{tr}
\end{equation}

As for $E=-mc^{2}\cos \theta $, the existence of a bound-state solution
requires $|v_{0}|>mc^{2}$ so that, defining $\tilde{\phi}^{\mathrm{T}%
}=\left( 1\text{ }i\cot \theta \right) $ for $v_{0}>+mc^{2}$, and $\tilde{%
\phi}^{\mathrm{T}}=\left( 0\text{ }1\right) $ for $v_{0}<-mc^{2}$, the
eigenspinor behaves asymptotically as $\phi \sim \tilde{\phi}f$ with%
\begin{equation}
f=\exp \left\{ -\frac{\sin \theta }{\hbar c}\left[ |v_{0}|+mc^{2}\text{sgn}%
\left( v_{0}x\right) \right] |x|\right\}  \label{efe}
\end{equation}

\section{The double-step potential}

Now we assume a scalar potential in the form%
\begin{equation}
V_{s}=v_{0}\left[ \Theta \left( x-a\right) -\Theta \left( -x-a\right) \right]
\end{equation}%
with $v_{0}$ and $a$ defined to be real numbers ($a>0$) and $\Theta \left(
x\right) $ is the Heaviside step function. Notice that $V_{s}$ is invariant
under a coincidental change of the signs of $v_{0}$ and $x$, and that as $%
a\rightarrow 0$ the double-step approximates the sign potential already
considered in Ref. \cite{aop1}.

Our problem is to solve the set of equations (\ref{6a})-(\ref{6b}) for $\phi
$ and to determine the allowed energies for both classes of solutions
sketched in Sec. 2.

\subsection{The case $E\neq -mc^{2}\cos \protect\theta $}

For our model,
\begin{equation}
V_{\mathtt{eff}}=\left( V_{1}+V_{2}\right) \Theta \left( x-a\right) +\left(
V_{1}-V_{2}\right) \Theta \left( -x-a\right) -\hbar \,\text{sgn}\left(
v_{0}\right) \sqrt{\frac{V_{1}}{2}}\left[ \delta \left( x-a\right) +\delta
\left( x+a\right) \right]  \label{pot}
\end{equation}%
where $\delta \left( x\right) =d\Theta \left( x\right) /dx$ is the Dirac
delta function and the following abbreviations have been used:
\begin{subequations}
\begin{eqnarray}
V_{1} &=&\,\frac{v_{0}^{2}\sin ^{2}\theta }{2c^{2}}  \label{par1} \\
&&  \notag \\
V_{2} &=&v_{0}\,\frac{E\cos \theta +mc^{2}}{c^{2}}  \label{par2}
\end{eqnarray}%
For $V_{1}=0$, $V_{\mathtt{eff}}$\ is an ascendant (a descendant) double
step if $V_{2}>0$ ($V_{2}<0$). For $V_{1}\neq 0$, the \textquotedblleft
effective potential\textquotedblright\ includes attractive (repulsive) delta
functions at $x=\pm a$ if $v_{0}>0$ ($v_{0}<0$). It has plateaus given by $%
V_{1}\pm V_{2}$ for $x\gtrless \pm a$ ($V_{\mathtt{eff}}=0$ for $|x|<a$) and
so we can consider scattering states for fermions coming from the left with $%
E_{\mathtt{eff}}>V_{1}-V_{2}$. Due to the discontinuities (for $V_{1}=0)$ or
singularities (for $V_{1}\neq 0)$ of $V_{\mathtt{eff}}$ at $x=\pm a$ one
should expect resonant transmission for certain values of $E_{\mathtt{eff}}$%
. Furthermore, for $V_{2}>V_{1}$ and $E_{\mathtt{eff}}<0$ one should expect
nonprogressive waves in the region $|x|<a$ and no transmission whereas the
transmission is ubiquitous for $V_{2}\leq 0$. Bound-state solutions for $%
|V_{2}|<V_{1}$ with $0<E_{\mathtt{eff}}<V_{1}-|V_{2}|$ should be expected
even if $v_{0}<0$ (delta functions apart, $\phi _{+}$ is insensitive to
simultaneous changes in the signs of $v_{0}$ and $x$). It is instructive to
note that $V_{1}$ tends to vanish as $|v_{0}|/mc^{2}\rightarrow 0$ so that
the effective potential becomes the double-step potential in a
nonrelativistic scheme.

We demand that $\phi _{+}$ be continuous at $x=\pm a$ , that is
\end{subequations}
\begin{equation}
\underset{\varepsilon \rightarrow 0}{\lim }\left( \left. \phi
_{+}\right\vert _{x=\pm a+\varepsilon }-\left. \phi _{+}\right\vert _{x=\pm
a-\varepsilon }\right) =0  \label{cont}
\end{equation}%
Otherwise, the differential equation for $\phi _{+}$ would contain
derivative of $\delta $-functions. Effects on $d\phi _{+}/dx$ in the
neighbourhood of $\ x=\pm a$ can be evaluated by integrating the
differential equation for $\phi _{+}$ from $\pm a-\varepsilon $ to $\pm
a+\varepsilon $ and taking the limit $\varepsilon \rightarrow 0$. The
connection formulas for $d\phi _{+}/dx$ can be summarized as%
\begin{equation}
\underset{\varepsilon \rightarrow 0}{\lim }\left( \left. \frac{d\phi _{+}}{dx%
}\right\vert _{x=\pm a+\varepsilon }-\left. \frac{d\phi _{+}}{dx}\right\vert
_{x=\pm a-\varepsilon }\right) =-\frac{v_{0}\sin \theta }{\hbar c}\left.
\phi _{+}\right\vert _{x=\pm a}  \label{discont}
\end{equation}

\subsubsection{Scattering states}

We turn our attention to scattering states for fermions coming from the left
as sketched in Sec. 3. The upper component for scattering states on the
entire space is written as
\begin{eqnarray}
\phi _{+} &=&[1-\Theta \left( x+a\right) ]\left( A_{+}e^{+i\zeta
_{-}x/a}+A_{-}e^{-i\zeta _{-}x/a}\right)  \notag \\
&&  \notag \\
&&+\left[ \Theta \left( x+a\right) -\Theta \left( x-a\right) \right] \left(
C_{+}e^{+i\xi x/a}+C_{-}e^{-i\xi x/a}\right)  \label{phii} \\
&&  \notag \\
&&+\Theta \left( x-a\right) B_{\pm }e^{\pm i\zeta _{+}x/a}  \notag
\end{eqnarray}%
where

\begin{subequations}
\label{def}
\begin{eqnarray}
\zeta _{\pm } &=&ak_{\pm } \\
&&  \notag \\
\xi &=&\frac{a}{\hbar c}\sqrt{E^{2}-m^{2}c^{4}}
\end{eqnarray}%
It is instructive to note that these quantities satisfy the constraint
\end{subequations}
\begin{equation}
\zeta _{-}^{2}+\zeta _{+}^{2}=2\left( \xi ^{2}-v^{2}\right)  \label{vin}
\end{equation}%
where%
\begin{equation}
v=\frac{av_{0}\sin \theta }{\hbar c}
\end{equation}%
With $\phi _{+}$ given by (\ref{phii}) and for $E>-mc^{2}\cos \theta $,
conditions (\ref{cont}) and (\ref{discont}) imply into the relative
amplitudes
\begin{subequations}
\label{amp}
\begin{eqnarray}
\frac{A_{-}}{A_{+}} &=&e^{-2i\zeta _{-}}\frac{\left( \zeta _{-}-\zeta
_{+}\right) \mu +i\sigma _{-}}{\left( \zeta _{-}+\zeta _{+}\right) \mu
-i\sigma _{+}}  \label{amp1} \\
&&  \notag \\
\frac{B_{+}}{A_{+}} &=&e^{-i\left( \zeta _{-}+\zeta _{+}\right) }\zeta _{-}%
\frac{2\xi }{\left( \zeta _{-}+\zeta _{+}\right) \mu -i\sigma _{+}}
\label{amp2} \\
&&  \notag \\
\frac{C_{\pm }}{A_{+}} &=&e^{-i\zeta _{-}}\zeta _{-}\frac{\eta _{\pm }}{%
\left( \zeta _{-}+\zeta _{+}\right) \mu -i\sigma _{+}}  \label{amp4}
\end{eqnarray}%
where we have set
\end{subequations}
\begin{subequations}
\label{mumu}
\begin{eqnarray}
\mu &=&\xi \cos 2\xi -v\sin 2\xi  \label{mu2} \\
&&  \notag \\
\sigma _{\pm } &=&2\xi v\cos 2\xi +\frac{\left( \zeta _{-}\pm \zeta
_{+}\right) ^{2}}{2}\sin 2\xi  \label{mu3} \\
&&  \notag \\
\eta _{\pm } &=&\left( \xi \pm \zeta _{+}\right) \cos \xi -v\sin \xi \mp i
\left[ v\cos \xi +\left( \xi \pm \zeta _{+}\right) \sin \xi \right]
\label{mu4}
\end{eqnarray}%
For $E<-mc^{2}\cos \theta $, the amplitudes can be obtained by taking the
complex conjugate of\ the right-hand side of (\ref{amp}) and exchanging the
signs of the subscripts of the amplitudes. It follows that
\end{subequations}
\begin{equation}
T=T_{\gtrless }=\frac{32\xi ^{2}\zeta _{-}\text{Re}\zeta _{+}}{c_{1}\cos
4\xi +c_{2}}
\end{equation}%
where
\begin{subequations}
\label{c8}
\begin{eqnarray}
c_{1} &=&\left( \zeta _{-}^{2}-\zeta _{+}^{2}\right) ^{2}-16\xi ^{2}v^{2} \\
&&  \notag \\
c_{2} &=&8\xi ^{2}\left( \zeta _{-}+\zeta _{+}\right) ^{2}-c_{1}
\end{eqnarray}%
taking no regard if $E>-mc^{2}\cos \theta $ or $E<-mc^{2}\cos \theta $.
Because we chose fermions coming from the left, the transmission coefficient
is not invariant under the change of $v_{0}$ by $-v_{0}$, however, this
symmetry is exact when $T\neq 0$. The transmission coefficient does not
depends on the sign of $E$ in the case of a pure scalar coupling. It is true
that $\xi $ imaginary makes $\zeta _{-}$ imaginary. Naturally, the
transmission does not exist neither does the incidence if $\xi $ is small
enough to make $\zeta _{-}$ imaginary. The transmission is possible only if $%
\zeta _{-}$ and $\zeta _{+}$ are real quantities and this fact imposes a
cutoff on $\xi $. As a function of $\xi $, the transmission coefficient
rises from zero for a certain $\xi $, oscillates between extreme values and
approaches $T=1$ as $\xi \rightarrow \infty $. This is in contrast to the
transmission coefficient for the sign potential \cite{aop1}, where no
oscillation exists. The maxima of $T$ occur when
\end{subequations}
\begin{equation}
\xi _{n}=\left( n-\frac{1}{2}\right) \frac{\pi }{2},\quad n_{\min
}<n=1,2,3,\ldots
\end{equation}%
and they are osculated by the function
\begin{equation}
T_{\max }=\frac{4\xi ^{2}}{\left( \zeta _{-}+\zeta _{+}\right) ^{2}}\frac{%
4\zeta _{-}\text{Re}\,\zeta _{+}}{\left( \zeta _{-}+\zeta _{+}\right)
^{2}+\left( 2v\right) ^{2}}
\end{equation}%
$n_{\min }$ is associated with the cutoff on $\xi $. On the other hand, the
minima of $T$ occur when $\xi =\left( n+1\right) \pi /2$, and they are
osculated by the function%
\begin{equation}
T_{\min }=\frac{4\zeta _{-}\text{Re}\,\zeta _{+}}{\left( \zeta _{-}+\zeta
_{+}\right) ^{2}+\left( 2v\right) ^{2}}
\end{equation}%
\ The maxima (or minima) are regularly separated from each other by $\xi
=\pi /2$ and seen as a function of $E$ \ they are more scarce the small $a$.
As a matter of fact, as $a\rightarrow 0$ one finds the transmission
coefficient for the sign step potential \cite{aop1}: $T\simeq T_{\min }$. It
is also worthwhile to note that the oscillations in the transmission
coefficient only manifest in a nonrelativistic scheme if $a\gg \lambda _{C}$%
. Fig. 1 shows the transmission coefficient as a function of $\xi $ for $%
v_{0}/mc^{2}=1$, $a/\lambda _{C}=1/2$ and $\theta =3\pi /8$.

\subsection{Bound states}

Following the previous qualitative considerations, we discuss the existence
of bound states. One way to identify the possible bound-state solutions is
to look for the poles of the transmission amplitude in the complex $\zeta $%
-plane. If one considers the transmission amplitude $t_{\pm }$ in (\ref{t})
as a function of the complex variables $\zeta _{\pm }$ one sees that bound
states would be obtained by the poles along the imaginary axis of the
complex $\zeta $-plane. These poles require $A_{\pm }=0$ and $B_{\pm }\neq 0$%
, corresponding to $E\gtrless -mc^{2}\cos \theta $. Furthermore,
square-integrability of $\phi _{+}$ demands $\zeta _{-}=\pm i|\zeta _{-}|$
and $\zeta _{+}=\pm i|\zeta _{+}|$. Therefore, the bound states would occur
for $\Sigma =\sigma _{+}/\mu $, even if $\xi =0$. Here $\Sigma =|\zeta
_{-}|+|\zeta _{+}|$. Hence,%
\begin{equation}
\Sigma =2v  \label{cq1}
\end{equation}%
or%
\begin{equation}
\Sigma =-2\xi \cot 2\xi  \label{cq2}
\end{equation}%
Eqs. (\ref{cq1}) and (\ref{cq2}) would determinate the energies of the bound
states. Notice that any possible solution demands $\Sigma >0$.

It is easy to see that (\ref{cq1}) is impossible for $v<0$. Eq. (\ref{cq1})
has not even solution for $v>0$. In fact, squaring (\ref{cq1}) results in
the form of a second-degree algebraic equation%
\begin{equation}
E^{2}+2mc^{2}E\cos \theta \,+m^{2}c^{4}\cos ^{2}\theta =0
\end{equation}%
which presents just one solution for $v>0$, viz. $E=-mc^{2}\cos \theta $,
without regard to $a$. Evidently, it is not a proper solution of the problem.

The remaining quantization condition, Eq. (\ref{cq2}), has no solution if $%
\xi $ is imaginary because its right-hand side would be a negative number.
For $\xi \in
%TCIMACRO{\U{211d} }%
%BeginExpansion
\mathbb{R}
%EndExpansion
$ ($|E|>mc^{2}$), though, it dictates that the spectrum depends on the
mixing angle and that it is symmetrical about $E=0$ in the case of a pure
scalar coupling. The sign of $v_{0}$ does not effect the spectrum but $a$
does. In the case of a massless fermion the spectrum does not change when
the mixing angle changes from $\pi /2-\varepsilon $ to $\pi /2+\varepsilon $%
. From (\ref{def}) one sees that $\zeta _{-}$ and $\zeta _{+}$ are imaginary
numbers only when $|v_{0}|>mc^{2}$ so that those solutions only survive in a
relativistic regime. Eq. (\ref{cq2}) also dictates that $\xi \cot 2\xi <0$
so that $\xi >\pi /4$. In addition, Eq. (\ref{vin}) becomes $|\zeta
_{-}|^{2}+|\zeta _{+}|^{2}=2\left( v^{2}-\xi ^{2}\right) $ for bound states
and so $\xi <|v|$. In this way we have to search for solutions in the
interval $\pi /4<\xi <|v|$, corresponding to%
\begin{equation}
\left( \frac{\pi \hbar c}{4a}\right) ^{2}<E^{2}-m^{2}c^{4}<(v_{0}\sin \theta
)^{2}\,
\end{equation}%
At once, we can state that the possible spectrum for bound states calls for $%
\sin \theta \neq 0$ and a minimum value for $a|v_{0}|\sin \theta $, namely
\begin{equation}
a|v_{0}|\sin \theta >\frac{\pi \hbar c}{4}  \label{Cond}
\end{equation}

The graphical method for $\xi \in
%TCIMACRO{\U{211d} }%
%BeginExpansion
\mathbb{R}
%EndExpansion
$ is illustrated in Fig. 2. The solutions for bound states are given by the
intersection of the curves represented by $\Sigma $ with the curve
represented by $-2\xi \cot 2\xi $. Seen as a function of $\xi $, $\Sigma $
is a two-branch decreasing function which begins with $\Sigma \simeq 2a\sin
\theta \sqrt{|v_{0}|\left( |v_{0}|+mc^{2}\right) }/\hbar c$ at $\xi =0$ and
ends for some $\xi >0$. The branch of solutions with $E<-mc^{2}\cos \theta $
($E>-mc^{2}\cos \theta $) is more favoured for $\theta $ $<\pi /2$ ($\theta $
$>\pi /2$). Above of critical values of $|v_{0}|$, $a$ and $\sin \theta $
there will be a finite sequence of bound states with%
\begin{equation}
\left( n-\frac{1}{2}\right) \frac{\pi }{2}<\xi _{n}<n\frac{\pi }{2},\quad
n=1,2,3,\ldots <\frac{2|v|}{\pi }
\end{equation}%
Hence,%
\begin{equation}
\left[ \left( n-\frac{1}{2}\right) \frac{\pi \hbar c}{2a}\right]
^{2}<E^{2}-m^{2}c^{4}<\left( n\frac{\pi \hbar c}{2a}\right) ^{2}
\end{equation}%
with%
\begin{equation}
a|v_{0}|\sin \theta >\left( n-\frac{1}{2}\right) \frac{\pi \hbar c}{2}
\label{cond}
\end{equation}%
and $|v_{0}|>mc^{2}$. In point of fact, from (\ref{cond}) one can see that (%
\ref{Cond}) is the condition that there is at least one bound-state solution
and that the number of possible bound states depends on the size of $%
a|v_{0}|\sin \theta $. The number of possible bound states is determined by
the maximum value of $n$ satisfying inequality (\ref{cond}). Numerical
solutions for the eigenenergies corresponding to $n=1$ as a function of $%
v_{0}/mc^{2}$ are shown in Fig. 3 for different values of $a$ and $\theta $.
The case of a massless fermion, as already discussed before with fulcrum on
the charge-conjugation and the chiral-conjugation operations, presents a
spectrum symmetrical about $E=0$ and seen as a function of $\theta $
exhibits an additional symmetry about $\theta =\pi /2$.

Exploiting (\ref{cq2}) and using the relative amplitudes (\ref{amp}) for $%
E>-mc^{2}\cos \theta $, we can write
\begin{subequations}
\label{Amp2}
\begin{eqnarray}
\frac{B_{+}}{A_{-}} &=&-e^{-(|\zeta _{-}|-|\zeta _{+}|)}\frac{2\xi }{(|\zeta
_{-}|-|\zeta _{+}|+2v)\sin \xi }  \label{Amp2a} \\
&&  \notag \\
\frac{C_{\pm }}{A_{-}} &=&-e^{-|\zeta _{-}|}\frac{\eta _{\pm }}{(|\zeta
_{-}|-|\zeta _{+}|+2v)\sin \xi }  \label{Amp2b}
\end{eqnarray}%
Note that $\eta _{\pm }=\xi \cos \xi +\left( |\zeta _{+}|-v\right) \sin \xi
\pm i\left[ \left( |\zeta _{+}|-v\right) \cos \xi -\xi \sin \xi \right] $ in
such a way that $\eta _{\pm }^{\ast }=\eta _{\mp }$ and $C_{\pm }^{\ast
}=C_{\mp }$. Therefore, as before, the relative amplitudes for $%
E<-mc^{2}\cos \theta $ can be obtained by taking the complex conjugate of\
the right-hand side of (\ref{amp2}) and exchanging the signs of the
subscripts of the amplitudes. Fig. 4 shows the normalized position
probability density for a massive fermion for the Sturm-Liouville solution
with $n=1$, $v_{0}/mc^{2}=5$, $a/\lambda _{C}=1$ and $\theta =3\pi /8$.

\subsection{The case $E=-mc^{2}\cos \protect\theta $}

In this case, one finds
\end{subequations}
\begin{equation}
v\left( x\right) =v_{0}\left[ a+\left( |x|-a\right) \Theta \left(
|x|-a\right) \right]  \label{vx}
\end{equation}%
As commented before, there is no solution for $\sin \theta =0$. The
normalizable solution for $\sin \theta \neq 0$ requires $|v_{0}|>mc^{2}$
with $\phi =N_{\gtrless }\,\tilde{\phi}f$. Here,
\begin{equation}
f=\exp \left( -\frac{\alpha }{2a}\left\{ r\left[ |x|\Theta \left(
|x|-a\right) +a\Theta \left( -|x|+a\right) \right] +x\,\text{\textrm{sgn}}%
\left( v_{0}\right) \right\} \right)
\end{equation}%
where%
\begin{equation}
\alpha =\frac{2a\sin \theta }{\lambda _{C}},\quad \lambda _{C}=\frac{\hbar }{%
mc},\quad r=\frac{|v_{0}|}{mc^{2}}
\end{equation}%
The normalization condition $\int_{-\infty }^{+\infty }dx\,\left( |\phi
_{+}|^{2}+|\phi _{-}|^{2}\right) =1$ allows one determine $N_{\gtrless }$.
One finds the position probability density to be%
\begin{equation}
|\phi |^{2}=\frac{\alpha }{2a}\frac{r^{2}-1}{r}\frac{e^{r\alpha }}{\cosh
\alpha +r\sinh \alpha }|f|^{2}  \label{den}
\end{equation}%
Therefore, a massive fermion tends to concentrate at the left (right) region
when $v_{0}>0$ ($v_{0}<0$), and tends to avoid the origin more and more as $%
\sin \theta $ decreases. A massless fermion has a position probability
density symmetric around the origin (though $r\rightarrow \infty $ and $%
\alpha \rightarrow 0$ as $m\rightarrow 0$, $r\alpha \rightarrow $ constant).
One can see that the best localization occurs for a pure scalar coupling. In
fact, the fermion becomes delocalized as $\sin \theta $ decreases. From%
\begin{equation}
\lim_{a\rightarrow 0}f=\exp \left\{ -\frac{\sin \theta }{\lambda _{C}}\left[
r+\text{sgn}\left( v_{0}x\right) \right] |x|\right\}
\end{equation}%
one recovers the value for $\phi $ in the case of the sign potential (at
small $a$) as in Ref. \cite{aop1}. Fig. 5 illustrates the position
probability density for a massive fermion with $r=5$ ($v_{0}>0$), $a/\lambda
_{C}=1$ and $\theta =3\pi /8$.

The expectation value of $x$ is given by%
\begin{equation}
<x>=-\text{sgn}\left( v_{0}\right) \frac{a}{\left( r^{2}-1\right) \alpha }%
\frac{c_{3}\cosh \alpha +c_{4}\sinh \alpha }{\cosh \alpha +r\sinh \alpha }
\label{espec1}
\end{equation}%
and the fermion is confined within an interval $\Delta x=\sqrt{%
<x^{2}>-<x>^{2}}$ given by%
\begin{equation}
\Delta x=\frac{a}{\sqrt{2}\left( r^{2}-1\right) \alpha }\frac{\sqrt{%
c_{5}\cosh 2\alpha +c_{6}\sinh 2\alpha +c_{7}}}{\cosh \alpha +r\sinh \alpha }
\label{dx}
\end{equation}%
where
\begin{subequations}
\label{c1}
\begin{eqnarray}
c_{3} &=&2+r\left( r^{2}-1\right) \alpha ,\quad c_{4}=\left( r^{2}-1\right)
\alpha -r\left( r^{2}-3\right) \\
&&  \notag \\
c_{5} &=&r^{6}+5r^{2}+2,\quad c_{6}=2r\left( r^{4}+3\right) \\
&&  \notag \\
c_{7} &=&2-r^{2}-r^{6}-2\alpha ^{2}\left( r^{2}-1\right) ^{3}-4r\alpha
\left( r^{2}-1\right) ^{2}
\end{eqnarray}%
Again one can see that the fermion becomes delocalized as $\sin \theta $
decreases and that the best localization occurs for a pure scalar coupling.
More than this, $|<x>|\rightarrow \infty $ and $\Delta x\rightarrow \infty $
as $r\rightarrow 1$. The values for $<x>$ and $\Delta x$ as $\alpha
\rightarrow 0$ (either in the case of $\sin \theta \rightarrow 0$ or $%
m\rightarrow 0$ or in the case of the sign potential as in Ref. \cite{aop1})
are given by
\end{subequations}
\begin{subequations}
\begin{eqnarray}
&<x>\rightarrow -\text{sgn}\left( v_{0}\right) \frac{2a}{\left(
r^{2}-1\right) \alpha }  \label{g1a} \\
&&  \notag \\
\Delta x \rightarrow &\frac{2a\sqrt{r^{2}+1}}{\sqrt{2}\left( r^{2}-1\right)
\alpha }  \label{g1b}
\end{eqnarray}%
The formula (\ref{g1b}) shows that $\Delta x$ decreases monotonically as $%
|v_{0}|$ increases. If $\Delta x$ reduces its extension then $\Delta p$ will
must expand, in consonance with the Heisenberg uncertainty principle.
Nevertheless, the maximum uncertainty in the momentum is comparable with $mc$
requiring that is impossible to localize a fermion in a region of space less
than or comparable with half of its Compton wavelength (see, for example,
\cite{gre}). This impasse can be broken by resorting to the concepts of
effective mass and effective Compton wavelength. Indeed, if one defines an
effective mass as $m_{\mathtt{eff}}=m\sqrt{r^{2}+1}$ and an effective
Compton wavelength $\lambda _{\mathtt{eff}}=\hbar /\left( m_{\mathtt{eff}%
}c\right) $, one will find
\end{subequations}
\begin{equation}
\Delta x_{\min }=\frac{\lambda _{\mathtt{eff}}}{\sqrt{2}\sin \theta }\frac{%
r^{2}+1}{r^{2}-1}  \label{dx1}
\end{equation}%
It follows that the high localization of fermions, related to small values
of $a$ and strong coupling, never menaces the single-particle interpretation
of the Dirac theory even if the fermion is massless ($m_{\mathtt{eff}%
}=|v_{0}|/c^{2}$). This fact is convincing because the scalar coupling
exceeds the vector coupling, and so the conditions for Klein's paradox are
never reached. As a matter of fact, (\ref{g1b}) furnishes $\left( \Delta
x\right) _{\min }\simeq \lambda _{\mathtt{eff}}/(\sqrt{2}\sin \theta )$ for $%
|v_{0}|\gg mc^{2}$.

\section{Final remarks}

After reviewing the use of a continuous chiral transformation for solving
the Dirac equation in the background of scalar and vector potentials,
already applied to the sign potential in Ref. \cite{aop1} and to the smooth
step potential in Ref. \cite{aop2}, we have extended the methodology to the
double-step potential. A common characteristic of all those kink-like
potentials is the appearance of an intrinsically relativistic isolated
bound-state solution corresponding to the zero-mode solution of the massive
Jackiw-Rebbi model in the case of no vector coupling. A finite set of
bound-state solutions appears as poles of the transmission amplitude in a
strong coupling regime for the double-step potential but all of these
solutions coming from the Sturm-Liouville problem tend to disappear as the
double-step potential approximates the sign potential. It was also shown
that all the bound solutions, including the isolated solution, disappear
asymptotically as the magnitude of the scalar and vector coupling becomes
the same. Furthermore, we show that due to the sizeable mass gain from the
scalar background the high localization of the fermion in an extreme
relativistic regime is conformable to comply with the Heisenberg uncertainty
principle. Therefore, those bound states can be highly localized by very
strong potentials without any chance of spontaneous pair production. A very
distinctive feature of the double-step potential is the sequence of
transmission resonances not seen in the other kinds of kink potentials (like
$\mathrm{sgn}\,x$ in \cite{aop1} and $\mathrm{\tanh }\,x\!/\!\lambda $ in
\cite{aop2}).

The adiabatic method for studying the charge fractionization developed in
Ref. \cite{gol} depends on the background field and it is only reliable for
small spacial gradients \cite{nie}. The complete set of stationary solutions
with sharp discontinuities for the kink-like background field might be
useful for additional studies of charge fractionization.

\bigskip

\bigskip

\bigskip

\textbf{Acknowledgments}

A.S.C. thanks CNPq (Conselho Nacional de Desenvolvimento Cient\'{\i}fico e
Tecnol\'{o}gico) under Project 304743/2015-1 for support.

\newpage

\begin{figure}[th]
\begin{center}
\includegraphics[width=14cm, angle=0]{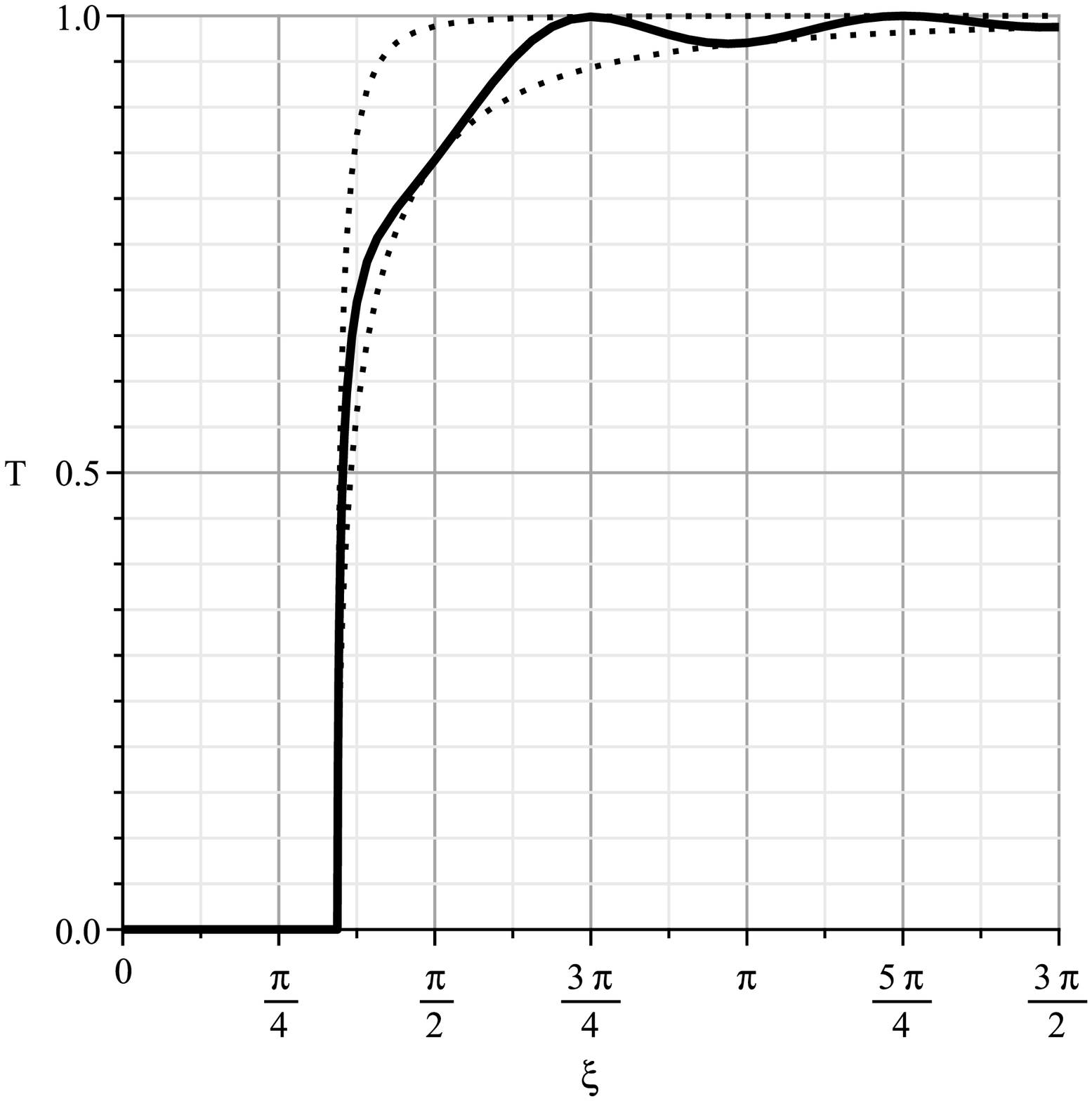}
\end{center}
\par
\vspace*{-0.1cm}
\caption{Transmission coefficient as a function of $\protect\xi $
(continuous line) for $v_{0}/mc^{2}=1$, $a/\protect\lambda _{C}=1/2$ and $%
\protect\theta =3\protect\pi /8$. The dotted lines osculate the minima and
maxima of the transmission coefficient.}
\end{figure}

\begin{figure}[th]
\begin{center}
\includegraphics[width=14cm, angle=0]{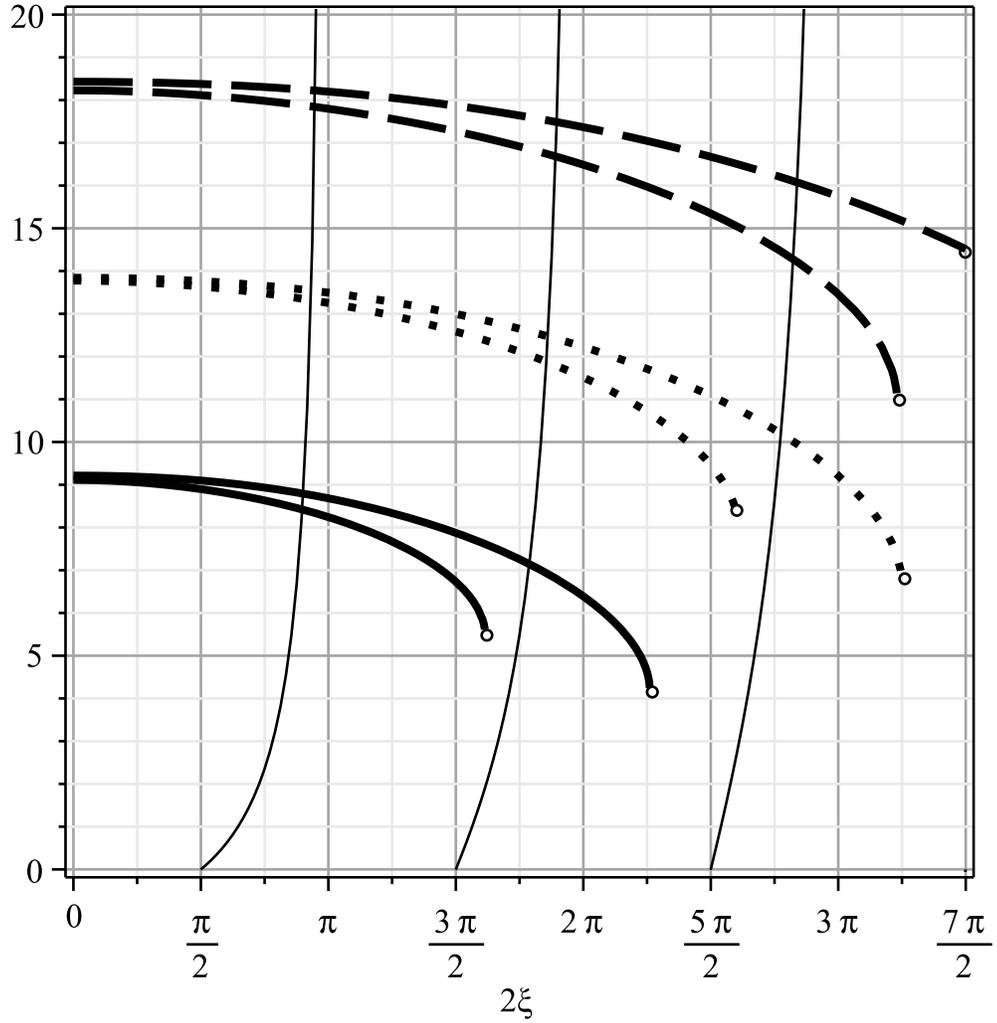}
\end{center}
\par
\vspace*{-0.1cm}
\caption{Graphical solution of $|\protect\zeta _{-}|+|\protect\zeta _{+}|=-2%
\protect\xi \cot 2\protect\xi $. The continuous tenuous line for $-2\protect%
\xi \cot 2\protect\xi $. The continuous, dotted and dashed lines for $|%
\protect\zeta _{-}|+|\protect\zeta _{+}|$ with $\protect\theta =3\protect\pi %
/8$, and $(v_{0}$,$a/\protect\lambda _{C})$ equal to $(10,1/2)$, $(15,1/2)$
and $(10,1)$, respectively. The higher (lower) curve of each pair $(v_{0}$,$%
a/\protect\lambda _{C})$ corresponds to $E<0$ ($E>0$).}
\end{figure}

\begin{figure}[th]
\begin{center}
\includegraphics[width=14cm, angle=0]{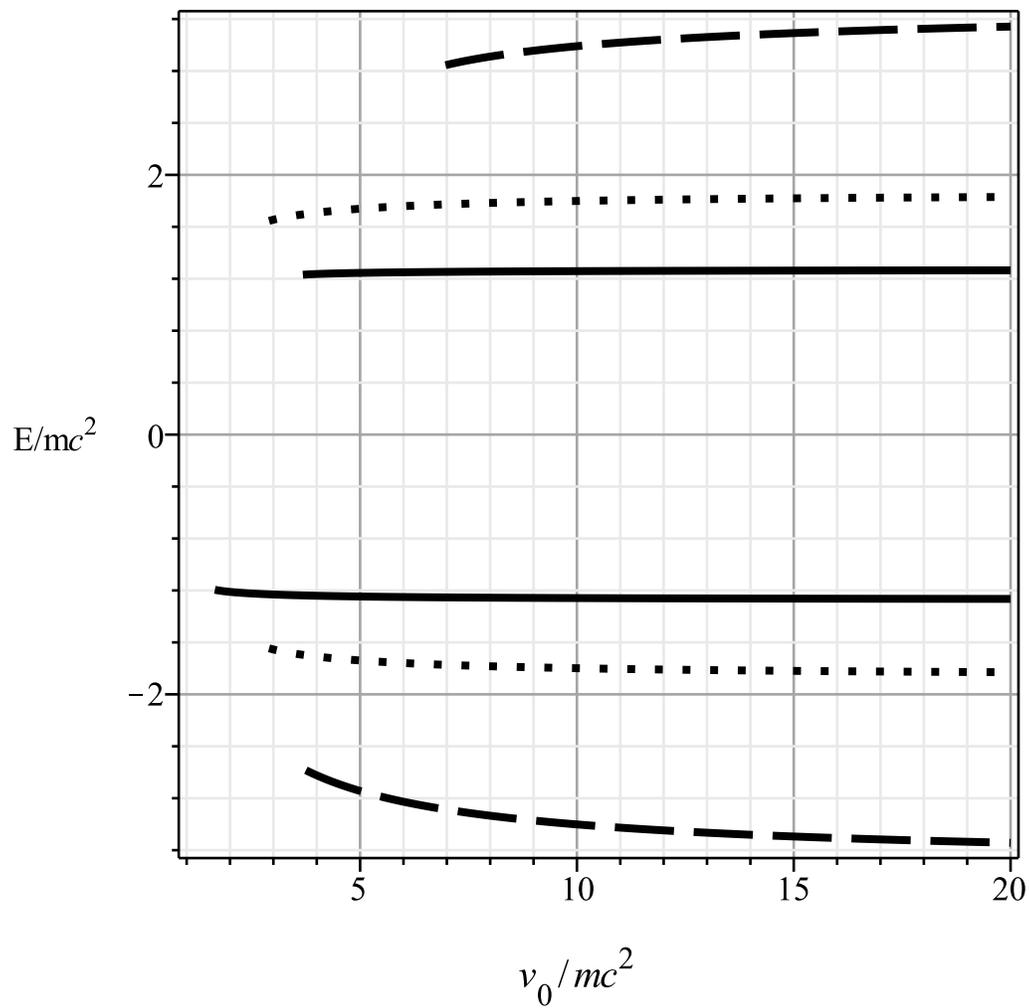}
\end{center}
\par
\vspace*{-0.1cm}
\caption{Energy levels as a function of $v_{0}/mc^{2}$ for $n=1$. The
continuous, dashed and dotted lines for $(a/\protect\lambda _{C},\protect%
\theta )$ equal to $(2,3\protect\pi /8)$, $(1/2,3\protect\pi /8)$ and $(1,%
\protect\pi /2)$, respectively.}
\end{figure}

\begin{figure}[th]
\begin{center}
\includegraphics[width=14cm, angle=0]{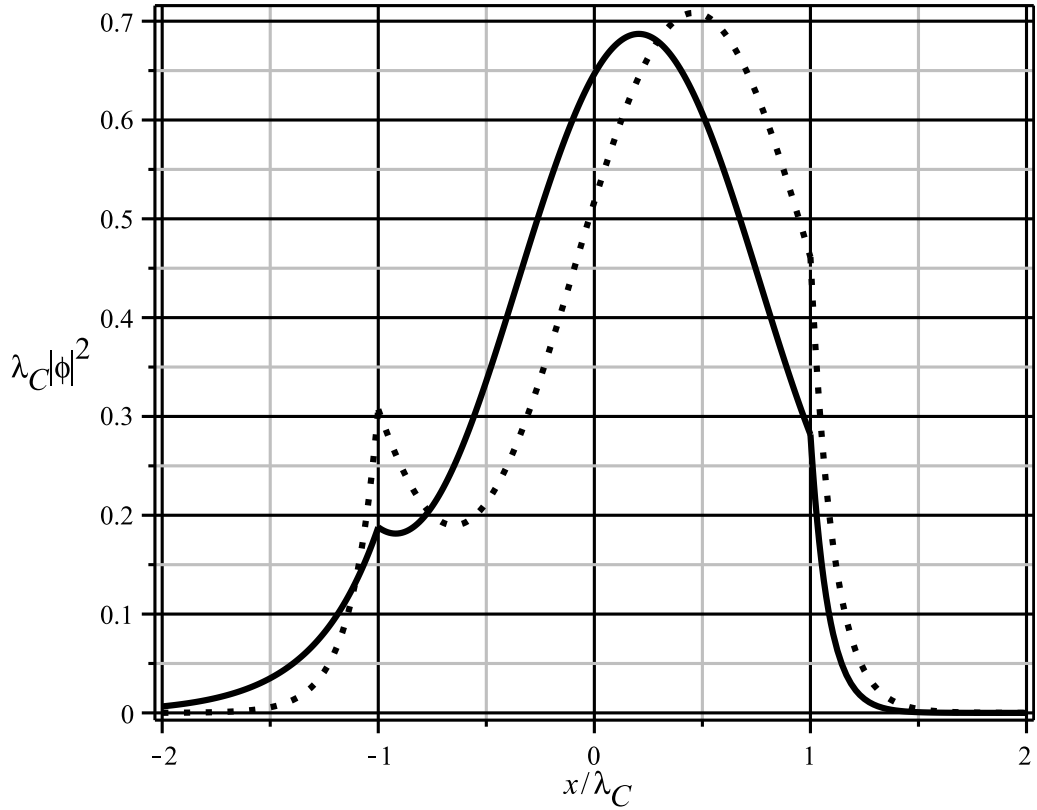}
\end{center}
\par
\vspace*{-0.1cm}
\caption{Position probability density for the Sturm-Liouville solution with $%
n=1$, $v_{0}/mc^{2}=5$, $a/\protect\lambda _{C}=1$ and $\protect\theta =3%
\protect\pi /8$. The continuous line for $E/mc^{2}=+1.7176$, and the dotted
line for $E/mc^{2}=-1.7321$. $\protect\lambda _{C}=\hbar /mc$ denotes the
Compton wavelength of the fermion.}
\end{figure}

\begin{figure}[th]
\begin{center}
\includegraphics[width=14cm, angle=0]{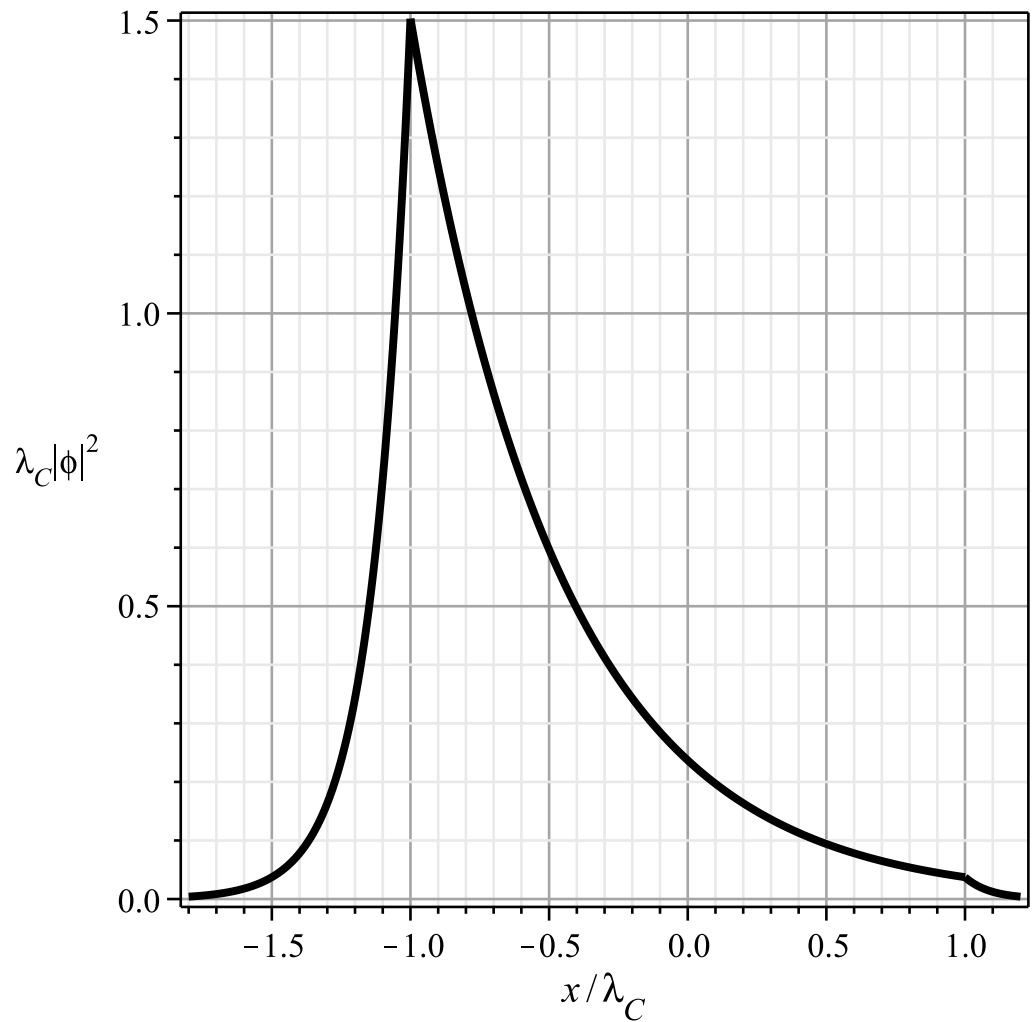}
\end{center}
\par
\vspace*{-0.1cm}
\caption{Position probability density for the isolated solution with $%
v_{0}/mc^{2}=5$, $a/\protect\lambda _{C}=1$ and $\protect\theta =3\protect%
\pi /8$.}
\end{figure}

\end{document}